\documentclass[submission,copyright,creativecommons]{eptcs}
\providecommand{\event}{FMAS 2023} 

\usepackage{iftex}
\usepackage{graphicx}
\usepackage{tikz}
\usepackage{url}
\usepackage{listings}
\usepackage{xspace}
\usepackage{color}
\tikzstyle{every node} =
[draw = none, fill = white, thin]
\tikzstyle{every edge} +=
[black, thick]

\tikzstyle{noall} =
[draw = none, fill = none]
\tikzstyle{nodraw} =
[draw = none, fill = white]
\tikzstyle{nofill} =
[draw = black, fill = none]

\tikzstyle{cnode} =
[circle, draw = black]
\tikzstyle{snode} =
[regular polygon, regular polygon sides = 4, draw = black]
\tikzstyle{lnode} =
[diamond, draw = black]

\newtheorem{definition}{Definition}
\newtheorem{example}{Example}
\newtheorem{theorem}{Theorem}

\renewcommand{\top}{\mathrm{tt}}
\renewcommand{\bot}{\mathrm{ff}}

\newcommand{\unk}{\mathrm{uu}}

\newcommand{\atmodels}[3]{{#1\models^{#2}_{#3}}}

\newcommand{\modelst}{\mathrel{\atmodels{}{3}{}}}

\newcommand{\tv}[2]{#1 \allowbreak\modelst\allowbreak #2}
\newcommand{\tvt}[2]{(\tv{#1}{#2}) = \top}
\newcommand{\tvf}[2]{(\tv{#1}{#2}) = \bot}

\def\qed{\hfill{\qedboxempty}      
	\ifdim\lastskip<\medskipamount \removelastskip\penalty55\medskip\fi}

\def\qedboxempty{\vbox{\hrule\hbox{\vrule\kern3pt
			\vbox{\kern3pt\kern3pt}\kern3pt\vrule}\hrule}}

\newenvironment{itemize*}%
{\begin{itemize}%
		\setlength{\itemsep}{1pt}%
		\setlength{\parskip}{0pt}}%
	{\end{itemize}}
\newenvironment{enumerate*}%
{\begin{enumerate}%
		\setlength{\itemsep}{1pt}%
		\setlength{\parskip}{0pt}}%
	{\end{enumerate}}

\lstnewenvironment{alghead}{\lstset{basicstyle={\normalsize},belowskip={-0.1cm},frame=single,language=Python,mathescape=true,tabsize=1,numbers=none}}{}
\lstnewenvironment{algbody}{\lstset{numbers=left,numberstyle=\scriptsize,numbersep=-4pt,basicstyle={\small},belowskip={-0.15cm},frame=single,language=Python,mathescape=true,tabsize=1}}{}

\newcommand{\ntool}{{3vLTL}\xspace}

\newcommand{\new}[1]{{\color{black}{#1}}}

\usepackage{amsmath,array}

\ifpdf
  \usepackage{underscore}         
  \usepackage[T1]{fontenc}        
\else
  \usepackage{breakurl}           
\fi

\title{\ntool: A Tool to Generate Automata \\ for Three-valued LTL}
\author{Francesco Belardinelli
\institute{Imperial College, London, United Kingdom}
\email{francesco.belardinelli@imperial.ac.uk}
\and
Angelo Ferrando 
\institute{University of Genoa, Genoa, Italy}
\email{angelo.ferrando@unige.it}
\and
Vadim Malvone
\institute{Telecom Paris, Paris, France}
\email{vadim.malvone@telecom-paris.fr}
}
\def\titlerunning{\ntool: A Tool to Generate Automata for Three-valued LTL}
\def\authorrunning{F. Belardinelli, A. Ferrando \& V. Malvone}

\begin{document}
\maketitle

\begin{abstract}
Multi-valued logics have a long tradition in the literature on system verification, including run-time verification. However, comparatively fewer model-checking tools have been developed for multi-valued specification languages. We present \ntool, a tool to generate B\"uchi automata from formulas in Linear-time Temporal Logic (LTL) interpreted on a three-valued semantics. Given an LTL formula, a set of atomic propositions as the alphabet for the automaton, and a truth value, our procedure generates a B\"uchi automaton that accepts all the words that assign the chosen truth value to the LTL formula. Given the particular type of the output of the tool, it can also be seamlessly processed by third-party libraries in a natural way. That is, the B\"uchi automaton can then be used in the context of formal verification to check whether an LTL formula is true, false, or undefined on a given model.
\end{abstract}

\section{Introduction}

Multi-valued logics have a long tradition in the literature on system verification, as demonstrated by various references \cite{Bruns+99a,Godefroid+03a,BallKupferman06,ShohamGrumberg04,HuthJS04,HuthP04}, and they play a crucial role in run-time verification as well~\cite{BauerLS06,bauer2007ugly}. Of particular interest are three-valued logics, including temporal extensions of Kleene's logic~\cite{Kleene+52a}, where the third value, in addition to true and false, is interpreted as "unknown" or "unspecified". Such semantics prove especially convenient when constructing smaller abstractions of complex reactive and distributed systems. These abstractions are typically approximations of the original model, containing strictly less information. Consequently, the challenge lies in finding the right trade-off between reducing complexity and minimizing information loss during the abstraction process.
In system verification, one of the most widely used temporal logics for specifying requirements is Linear-time Temporal Logic ($LTL$)~\cite{Pnu77}. The model checking problem for $LTL$ is typically addressed through automata-theoretic techniques~\cite{Baier2008}. Given a model $M$ of a transition system and an $LTL$ formula $\varphi$, the approach involves generating B\"uchi automata for both $M$ and the negation of $\varphi$. This allows us to determine whether $\varphi$ is satisfied in $M$ by examining whether the language accepted by the product of these two automata is empty.

Several tools are available now to generate B\"uchi automata from $LTL$ formulas. Notable examples include \cite{DBLP:conf/atva/Duret-LutzLFMRX16,DBLP:conf/cav/GastinO01}. However, to the best of our knowledge, no tool has yet been proposed to directly generate B\"uchi automata for multi-valued temporal logics.

{\bf Contribution.} In this paper we present \ntool, a tool to generate (generalized) B\"uchi automata from $LTL$ formulas interpreted on a three-valued semantics. Specifically, given an $LTL$ formula, a set of atomic propositions (representing the automaton alphabet), and a truth value (true, false or undefined), our procedure generates a B\"uchi automaton that accepts all the words that assign the chosen truth value to the input $LTL$ formula. Furthermore, \ntool has the functionality to process our output (i.e., the automaton) by third-party libraries in a natural way.
The present work is motivated by the use of three-valued logics in system verification. 
Indeed, our tool can be used in several works, such as \cite{konikowska1998three,godefroid2009ltl,timm2016parameterised,tzoref2006automatic}, to provide results for the verification of three-valued $LTL$ formulas.
Furthermore, our tool is already used in \cite{AIJ2022}. In this work, the authors present an abstraction-refinement method to partially solve the model checking of multi-agent systems under imperfect information and perfect recall strategies.
\new{Note that, a three-valued semantics becomes particularly significant in situations involving imperfect information, as the absence of information can lead to the emergence of a third value. This is particularly evident in autonomous and distributed systems, where a component may not have access to the complete system's information~\cite{DBLP:conf/sefm/FerrandoM22}.}

{\bf Related Work.} 
Concerning the three-valued automata technique for $LTL$ employed in this work, the most closely related approaches can be found in \cite{KupfermanL07,chechik2001model,Bruns+03a,DBLP:conf/forte/VijzelaarF17}. Notably, in \cite{Bruns+03a}, there is an exploration of a reduction from multi-valued to two-valued $LTL$, but it does not encompass automata-theoretic techniques. Conversely, in \cite{chechik2001model}, an automata-theoretic approach for general multi-valued $LTL$ is presented, following the tableau-based construction as outlined in \cite{Gerth+95a}; however, this work is more suitable for on-the-fly verification w.r.t. to our approach. 
In a different vein, \cite{KupfermanL07} delves into general multi-valued automata, defining lattices, deterministic and non-deterministic automata, as well as their extensions with B\"uchi acceptance conditions. As part of their theoretical findings, they introduce an automata construction for multi-valued $LTL$, though it lacks a clear explanation of states and transitions.
With respect to our work, in \cite{KupfermanL07}, the model checking is only briefly discussed, and their approach is tailored more toward multi-valued logics in a broader sense.

\begin{sloppypar}
To summarize, unlike \cite{KupfermanL07,chechik2001model,Bruns+03a}, our proposed approach makes minimal modifications to the automata-theoretic construction for two-valued $LTL$ \cite{Baier2008} and extends it to a three-valued interpretation. 
\end{sloppypar}

\section{Preliminaries}\label{sec:preliminaries}

In this part we present a three-valued semantics for Linear-time
Temporal Logic $LTL$ and recall the definition of generalized
non-deterministic B\"uchi automata.  To fix the notation, we assume
that $AP = \{q_1, q_2, \ldots \}$ is the set of {\em atomic
  propositions}, or simply atoms.  We denote the length of a tuple $t$
as $|t|$, and its $i$-th element as $t_i$.
For $i \leq |t|$, let $t_{\geq i}$ be the suffix $t_{i},\ldots,
t_{|t|}$ of $t$ starting at $t_i$ and $t_{\leq i}$ its prefix
$t_{1},\ldots, t_{i}$. Notice that we start enumerations with index 1.
\vspace*{-1em}
\paragraph{Models.} We begin by giving a formal definition of Transition Model~\cite{Baier2008}.

\begin{definition}[Transition Model] \label{def:kripke}
	Given a set $AP$ of atoms, a \emph{Transition Model} is a
        tuple $M = \langle S, s_0, \longrightarrow, V \rangle$ such that
(i) $S$ is a finite, non-empty set of {\em states}, with {\em initial state}  $s_0 \in S$;
%
(ii) $\longrightarrow \subseteq S \times S$ is a serial {\em transition relation}; 
		%
(iii) $V: S \times AP \rightarrow \{\top,\bot,\unk{}\}$ is the {\em three-valued labelling function}.
\end{definition}

A path $p \in S^{\omega}$ is an infinite sequence $s_1 s_2 s_3\dots$ of states where $s_{i} \longrightarrow s_{i+1}$, for each $i \geq 1$. 
\vspace*{-1em}
\paragraph{Syntax.} Here, we recall the syntax of $LTL$.
\begin{definition}[$LTL$] \label{def:LTL}
Formulas in $LTL$ are defined as
	follows, where $q \in AP$:
 \vspace*{-1em}
	\begin{eqnarray*}
		\varphi & ::= & q \mid \neg \varphi \mid \varphi \land \varphi \mid X \varphi \mid (\varphi U \varphi)
	\end{eqnarray*}
	
\end{definition}

The meaning of operators \emph{next} $X$ and \emph{until} $U$ is
standard \cite{Baier2008}. Operators \emph{release} $R$,
\emph{finally} $F$, and \emph{globally} $G$ can be introduced as
usual:
$\varphi R \psi
\equiv \neg(\neg \varphi U \neg \psi)$, $F \varphi \equiv \top\: U
\varphi$, $G \varphi \equiv \bot R \varphi$.
\vspace*{-1em}
\paragraph{Semantics.} Formally we define the three-valued semantics for $LTL$ as follows.
\begin{definition}[Satisfaction] \label{3satisfaction}
	The three-valued satisfaction relation $\modelst$ for a Transition Model $M$, path $p \in S^{\omega}$, atom $q \in AP$, $v \in \{\top, \bot \}$, and formula $\varphi$ is defined as follows:
	\begin{tabbing} 
		$((M, s) \modelst A \psi )= 
		\top$ \ \ \ \ \ \ \ \ \=  iff \ \ \ \= for all paths $p$ in $M$, $\tvt{(M, p)}{\psi }$\kill
		$((M, s) \modelst A \psi ) = 
		\bot$ \> iff \>  for some path $p$ in $M$, $\tvf{(M, p)}{\psi }$\\
		$((M, p) \modelst q )= v$ \> iff \> ${V}(p_1,q) = v$\\
		$((M, p) \modelst \neg \psi )= v$ \> iff \> $((M, p) \modelst \psi )= \neg v$\\
		$((M, p) \modelst \psi \land \psi') = \top$ \>iff\>  $((M, p) \modelst \psi) = \top$  and $((M, p) \modelst \psi') = \top$\\
		$((M, p) \modelst \psi \land \psi') = \bot$ \>iff\>  $((M, p) \modelst \psi) = \bot$  or  $((M, p) \modelst \psi') = \bot$\\
		$((M, p) \modelst X \psi) = v$ \> iff \> $((M, p_{\geq 2}) \modelst \psi) = v$\\
		$((M, p) \modelst \psi U \psi') = \top$  \> iff \> for some $k \geq 1$, $((M, p_{\geq k}) \modelst \psi') = \top$, and \\
		\> \> for all $j$,  $1 \leq j < k \Rightarrow ((M, p_{\geq j}) \modelst \psi) = \top$ \\
		$((M, p) \modelst \psi U \psi') = \bot$  \> iff \> for all $k \geq 1$, $((M, p_{\geq k}) \modelst \psi') = \bot$, or \\
		\> \>for some $j \geq 1$, 
		$((M, p_{\geq j}) \modelst \psi) = \bot$, and \\
		\> \> for all $j'$,  $1 \leq j' \leq j \Rightarrow ((M, p_{\geq j'}) \modelst \psi') = \bot$
	\end{tabbing}
	In all other cases the value of $\varphi$ is $\unk{}$.
\end{definition}

\paragraph{Generalized non-deterministic B\"uchi automaton.} Now, we recall the definition of the class of automa that we will use in our construction and in the tool.
\begin{definition}[GNBA]
	A \emph{generalized non-deterministic B\"uchi automaton} is a tuple $A = \langle
	Q,Q_0, \Sigma, \pi, \mathcal{F} \rangle$ where
	%
		(i) $Q$ is a finite set of {\em states} with $Q_0 \subseteq Q$ as the set of
		{\em initial states};
		(ii) $\Sigma$ is an {\em alphabet};
		(iii) $\pi: Q \times \Sigma \rightarrow 2^Q$ is the (non-deterministic) {\em transition function};
		(iv) $\mathcal{F}$ is a (possibly empty) set of subsets of $Q$, whose elements are called {\em acceptance sets}.
\end{definition}
%
Given an infinite run $\rho = q_0 q_1 q_2 \ldots \in Q^{\omega}$, let $\textit{Inf}(\rho)$ be the set of states $q$ for which there are infinitely many indices $i$ with $q = q_i$, that is, $q$ appears infinitely often in $\rho$.
Then, run $\rho$ is {\em accepting} if for each acceptance
set $F \in \mathcal{F}$, $\textit{Inf}(\rho) \cap F \neq \emptyset$, that is,  there are infinitely many indices $i$ in $\rho$
with $q_i \!\in\! F.$
The {\em accepted language} $L(A)$ of automaton $A$ consists of all infinite words
$w \in \Sigma^\omega$ for which there exists at least one accepting
run $\rho = q_0 q_1 q_2 \ldots \in Q^{\omega}$ such that for all $i \geq 0$, $q_{i+1} \in \pi(q_i, w_i)$.

\section{Automata Construction}\label{sec:theory}In this section we provide a slightly variant of the automata-theoretic approach to the
verification of the three-valued linear-time logic $LTL$ as proposed
in \cite{BelardinelliM20}. In particular, in what follows we generalize the construction in \cite{BelardinelliM20} for the truth values $\top$, $\bot$, and $\unk$.
%
%
\begin{definition}[Closure and Elementarity]\label{closure}  \label{consistent}
	The {\em closure} $cl(\psi)$ of an $LTL$ formula $\psi$ is the set
	consisting of all subformulas $\phi$ of
	$\psi$ and their negation $\neg\phi$.
	A set $B \subseteq cl(\psi)$ is {\em consistent}
	w.r.t.~propositional logic
	iff for all
	$\psi_1 \land \psi_2, \neg \phi \in
	cl(\psi)$:
	(i) $\psi_1 \land \psi_2 \in
	B$ iff $\psi_1 \in B$ and $\psi_2 \in B$;
	(ii) $\neg(\psi_1 \land \psi_2) \in
	B$ iff $\neg \psi_1 \in B$ or $\neg \psi_2 \in B$;
	%
	(iii) if $\phi \in
	B$ then $\neg\phi \not\in B$;
	(iv) $\neg \neg \phi \in B$ iff $\phi \in B$.
	%
	\label{locallyconsistent}
	Further, $B$ is {\em locally consistent}
	w.r.t.~the until 	operator
	iff for all $\psi_1 U \psi_2 \in cl(\psi)$:
	(i) if $\psi_2 \in B$ then $\psi_1 U \psi_2 \in B$;
	(ii) if $\neg (\psi_1 U \psi_2) \in B$ then $\neg \psi_2 \in B$;
	(iii) if $\psi_1 U \psi_2 \in B$ and $\psi_2 \not\in B$ then $\psi_1 \in B$;
	(iv) if  $\neg \psi_1, \neg \psi_2 \in B$, then $\neg (\psi_1 U \psi_2) \in B$.
	
	Finally, $B$ is \emph{elementary} iff it is both consistent
	and locally consistent.
\end{definition}

Note that, unlike the standard construction for two-valued
$LTL$ \cite{Baier2008}, we do not require elementary sets
to be maximal (i.e., either $\phi \in B$ or $\neg \phi \in B$), but we do require extra conditions $(ii)$ and $(iv)$ on
consistency, and $(ii)$ and $(iv)$ on local consistency. These extra conditions can be derived in the classic, two-valued semantics, but need to be assumed as primitive here.

%
Hereafter $Lit = AP \cup \{\neg q \mid q \in AP\}$ is the set of {\em
	literals}.
\begin{definition}[Automaton $A_{\psi,v}$]
	\label{aut}
	Let $\psi$ be a formula in $LTL$.
	We define the automaton
	$A_{\psi,v} = \langle Q, Q_0, \allowbreak 2^{Lit}, \pi,\mathcal{F} \rangle$, where $v \in \{\top, \bot, \unk \}$, as follows:
		%
		$Q$ is the set of all elementary sets $B \subseteq cl(\psi)$.
        if $v = \top$ then $Q_0 = \{B \in Q \mid \psi \in B \}$; else if $v = \bot$ then $Q_0 =\{B \in Q \mid \neg \psi \in B \}$; otherwise $Q_0 = \{B \in Q \mid \psi \not\in B \textit{ and } \neg \psi \not\in B \}$.
		%
		The transition relation $\pi$ is given by:
		let $A \subseteq Lit$.
		If $A \neq B \cap Lit$, then $\pi(B,A) = \emptyset$;
		otherwise $\pi(B,A)$ is the set of all elementary sets $B'$ of formulas 
		such that for every $X\phi, \psi_1 U \psi_2  \in cl(\psi)$:
		(i)     $X\phi \in B$ iff $\phi \in B'$;
		(ii)    $\neg X\phi \in B$ iff $\neg \phi \in B'$;
		(iii)     $\psi_1 U \psi_2 \in B$ iff $\psi_2 \in B$ or, $\psi_1 \in B$  and $\psi_1 U \psi_2 \in B'$;
		(iv)    $\neg (\psi_1 U \psi_2) \in B$ iff $\neg \psi_2 \in B$ and, $\neg \psi_1 \in B$ or $\neg (\psi_1 U \psi_2) \in B'$.
		$\mathcal{F} = \{ F_{\psi_1 U \psi_2}
		\mid \psi_1 U \psi_2 \in cl(\psi)\} \cup \{Q\}$, where
		$F_{\psi_1 U \psi_2} = \{B \in Q \mid \psi_1 U \psi_2 \in B \textit{ implies } \psi_2 \in B \textit{ and } \neg \psi_2 \in B  \textit{ implies } \neg(\psi_1 U \psi_2)  \in B \}$.
\end{definition}

\new{According to Def. \ref{aut}, the transition relation operates as follows: when the automaton reads a set $A$ of literals that do not exist in the current state, the transition remains undefined. However, if these literals are present in the state, the automaton proceeds to verify the enabled transitions based on the semantics of the $LTL$ operators.
It is worth noting that in Def. \ref{aut}, we must also specify conditions for negated formulas. This is necessary because elementary sets may not necessarily be maximal in this context.
}

We present an example of automaton for the next operator and truth value undefined. 

\begin{figure}[t]		\centering
	\scalebox{0.55}{
		    \tikzstyle{state}=[circle,draw,trans, minimum size=8mm]
    \tikzstyle{initstate}=[circle,draw,trans, minimum size=8mm, fill=yellow]
    \tikzstyle{prop}=[]
    \tikzstyle{trans}=[font=\footnotesize]

\tikzstyle{every node} =
[draw = none, fill = white, thin]
\tikzstyle{every edge} +=
[black, thick]

\tikzstyle{noall} =
[draw = none, fill = none]
\tikzstyle{nodraw} =
[draw = none, fill = white]
\tikzstyle{nofill} =
[draw = black, fill = none]

\tikzstyle{cnode} =
[circle, draw = black]
\tikzstyle{snode} =
[regular polygon, regular polygon sides = 4, draw = black]
\tikzstyle{lnode} =
[diamond, draw = black]

\begin{tikzpicture}
[node distance = 5em]
\node [initstate]
(S0)
{$\stackrel{B_2}{a}$};
\node [initstate]
(S20) [above of = S0, node distance = 9em]
{$\stackrel{B_1}{\emptyset}$};
\node [initstate]
(S21) [below of = S0, node distance = 9em]
{$\stackrel{B_3}{\neg a}$};
\node [state]
(S1)
[above right of = S0, node distance = 7em]
{$\stackrel{B_4}{Xa}$};
\node [state]
(S2)
[below right of = S0, node distance = 7em]
{$\stackrel{B_5}{\neg Xa}$};
\node []
(S4)
[below right of = S1, node distance = 7em]
{$\stackrel{}{}$};
\node [state]
(S5)
[right of = S1, node distance = 8em]
{$\stackrel{B_6}{\{a, Xa\}}$};
\node [state]
(S6)
[right of = S2, node distance = 8em]
{$\stackrel{B_7}{\{\neg a, Xa\}}$};
\node []
(S9)
[below right of = S5, node distance = 7em]
{$\stackrel{}{}$};
\node [state]
(S11)
[below right of = S9, node distance = 7em]
{$\stackrel{B_9}{\{\neg a, \neg Xa\}}$};
\node [state]
(S12)
[above right of = S9, node distance = 7em]
{$\stackrel{B_{8}}{\{a,\neg Xa\}}$};

\path[->]
(S0)  edge  [pos = 0.3]
node [above] {\footnotesize $\{a\}$}
(S1)
edge  [pos = 0.7]
node [above] {\footnotesize $\{a\}$}
(S2)
edge  [pos = 0.5]
node [above] {\footnotesize $\{a\}$}
(S20)

(S20) 
edge  [loop above]
node [left] {\footnotesize $\emptyset$}
()
edge  [bend left = 25, pos = 0.5]
node [above] {\footnotesize $\emptyset$}
(S1)
edge  [bend left = 18, pos = 0.8]
node [above] {\footnotesize $\emptyset$}
(S2)

(S21) 
edge  [bend left = 25, pos = 0.6]
node [above] {\footnotesize $\{\neg a\}$}
(S20)
edge  [bend left = 10, pos = 0.2]
node [above] {\footnotesize $\{\neg a\}$}
(S1)
edge  [pos = 0.3]
node [right] {\footnotesize $\{\neg a\}$}
(S2)

(S1)  
edge  [pos = 0.3]
node [right] {\footnotesize $\emptyset$}
(S5)
edge  [bend left = 60, pos = 0.5]
node [right] {\footnotesize $\emptyset$}
(S12)
edge  [bend right = 30, pos = 0.5]
node [above] {\footnotesize $\emptyset$}
(S0)

(S2)  edge  [pos = 0.3]
node [right] {\footnotesize $\emptyset$}
(S6)
edge  [bend right = 30, pos = 0.3]
node [right] {\footnotesize $\emptyset$}
(S11)
edge  [bend left = 45, pos = 0.6]
node [below] {\footnotesize $\emptyset$}
(S21)

(S5)  edge  [pos = 0.3]
node [right] {\footnotesize $\{a\}$}
(S12)
edge  [pos = 0.5]
node [right] {\footnotesize $\{a\}$}
(S0)
edge  [loop above, pos = 0.2]
node [left] {\footnotesize $\{a\}$}
()

(S6) edge  [pos = 0.5]
node [above] {\footnotesize $\{\neg a\}$}
(S5)
edge  [pos = 0.3]
node [right] {\footnotesize $\{\neg a\}$}
(S0)
edge  [pos = 0.5]
node [left] {\footnotesize $\{\neg a\}$}
(S12)

(S11)  edge  [pos = 0.85]
node [right] {\footnotesize $\{\neg a\}$}
(S6)
 edge  [bend left = 35, pos = 0.5]
node [right] {\footnotesize $\{\neg a\}$}
(S21)
edge  [loop below, pos = 0.5]
node [right] {\footnotesize $\{\neg a\}$}
()

(S12)  edge  [bend right = 30, pos = 0.5]
node [above] {\footnotesize $\{a\}$}
(S6)
edge  [bend left = 50, pos = 0.2]
node [above] {\footnotesize $\{a\}$}
(S21)
edge  [pos = 0.7]
node [above] {\footnotesize $\{a\}$}
(S11)

;
\end{tikzpicture}
	} 
	\caption{The automaton $A_{\psi, \unk}$ for formula $\psi = Xa$. Initial states are marked in yellow.} \label{next}
\end{figure}
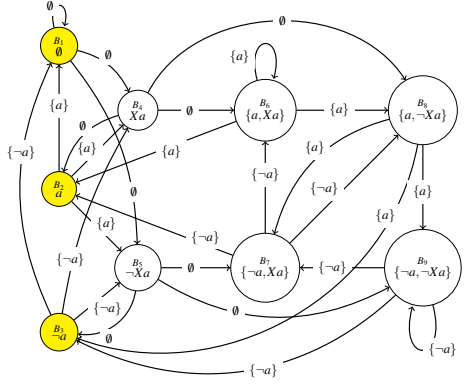

\begin{example}\label{example}
	Consider $\psi = Xa$. The GNBA $A_{\psi,\unk}$ in
	Fig.~\ref{next} is obtained as indicated in Def.~\ref{aut}.
	Namely, the state space $Q$ 
	consists of all
	elementary sets of formulas contained in $cl(\psi) = \{a,\neg
	a,X a,\neg X a \}$:
	$B_1 = \emptyset$, $B_2 =\{
	a\}$, $B_3 =\{ \neg a\}$, $B_4 =\{ X a\}$, $B_5 =\{ \neg Xa\}$, $B_6
	=\{ a, Xa\}$, $B_7 =\{ a, \neg Xa\}$, $B_8 =\{ \neg a, Xa\}$, $B_9
	=\{ \neg a, \neg Xa\}$.  The initial states of $A_{\psi,\unk}$ are the
	elementary sets $B \in Q$ with $\psi, \neg \psi \not \in B$. That is,
	$Q_0 = \{B_1,B_2,B_3\}$. The transitions are depicted in
	Fig.~\ref{next}.  The set $\mathcal{F}$ is $\{Q\}$ as $\psi$ does not
	contain until operators. Hence,
	every infinite run in $A_{\psi,\unk}$ is accepting.
\end{example}

Now, we provide a generalization of the main theoretical result proved in \cite{BelardinelliM20}.

\begin{theorem} \label{main}
	For every $LTL$ formula $\psi$ and truth value $v \in \{\top, \bot, \unk \}$,
	there exists a GNBA $A_{\psi, v}$
	(given as in Def.~\ref{aut}) 
	s.t.~$L(A_{\psi, v}) = Paths(\psi,v)$, where $Paths(\psi,v)$ is the set of paths $p \in (2^{Lit})^{\omega}$ such that $(p \models^3 \psi) = v$.
	%
	Moreover, the size of $A_{\psi,v}$
	is exponential in the size of $\psi$.
\end{theorem}


\definecolor{light-gray}{gray}{0.9}
\lstset{basicstyle=\ttfamily,
  basicstyle=\scriptsize,
  showstringspaces=false,
  commentstyle=\color{red},
  keywordstyle=\color{blue},
  backgroundcolor = \color{light-gray},
  frame=single,
  escapechar=\&
}

\section{Implementation}


\paragraph{Tool Architecture.}


The \ntool tool\footnote{\url{https://github.com/AngeloFerrando/3vLTL}}
developed for this paper aims at generating highly reusable
generalized non-deterministic B\"uchi automata (GNBA)
\cite{BelardinelliM20}. Hence, instead of generating only a
graphical result, \ntool produces a machine-readable file which can
be easily parsed by third-party tools and libraries as well.
From an engineering perspective, a pure graphical representation would
help the final user to visualise the generated automaton, but it
would not make it accessible for further evaluations.

\tikzset{every picture/.style={line width=0.75pt}} 

\begin{figure}[ht]
\centering

\scalebox{0.4}{

\begin{tikzpicture}[x=0.75pt,y=0.75pt,yscale=-1,xscale=1]

\draw   (185,61) -- (291,61) -- (291,120.8) -- (185,120.8) -- cycle ;
\draw   (368,71.8) -- (368,28.9) -- (375.31,19) -- (407,19) -- (407,71.8) -- cycle ;
\draw   (354,125) -- (424,125) -- (424,165) -- (354,165) -- cycle(418,131) -- (360,131) -- (360,159) -- (418,159) -- cycle ;
\draw    (74,37) .. controls (113.2,7.6) and (141.84,99.98) .. (180.61,75.45) ;
\draw [shift={(183,73.8)}, rotate = 143.13] [fill={rgb, 255:red, 0; green, 0; blue, 0 }  ][line width=0.08]  [draw opacity=0] (8.93,-4.29) -- (0,0) -- (8.93,4.29) -- cycle    ;
\draw    (90,82.8) .. controls (129.2,53.4) and (141.51,120.04) .. (179.63,94.48) ;
\draw [shift={(182,92.8)}, rotate = 143.13] [fill={rgb, 255:red, 0; green, 0; blue, 0 }  ][line width=0.08]  [draw opacity=0] (8.93,-4.29) -- (0,0) -- (8.93,4.29) -- cycle    ;
\draw    (88,145) .. controls (118.38,93.84) and (143,140.84) .. (181.62,114.5) ;
\draw [shift={(184,112.8)}, rotate = 143.13] [fill={rgb, 255:red, 0; green, 0; blue, 0 }  ][line width=0.08]  [draw opacity=0] (8.93,-4.29) -- (0,0) -- (8.93,4.29) -- cycle    ;
\draw    (292,75.8) .. controls (331.2,46.4) and (327.18,88.07) .. (364.65,61.51) ;
\draw [shift={(367,59.8)}, rotate = 143.13] [fill={rgb, 255:red, 0; green, 0; blue, 0 }  ][line width=0.08]  [draw opacity=0] (8.93,-4.29) -- (0,0) -- (8.93,4.29) -- cycle    ;
\draw    (291,108.8) .. controls (330.2,79.4) and (312.74,160.44) .. (349.67,135.46) ;
\draw [shift={(352,133.8)}, rotate = 143.13] [fill={rgb, 255:red, 0; green, 0; blue, 0 }  ][line width=0.08]  [draw opacity=0] (8.93,-4.29) -- (0,0) -- (8.93,4.29) -- cycle    ;
\draw  [fill={rgb, 255:red, 0; green, 0; blue, 0 }  ,fill opacity=1 ] (364,138) .. controls (364,135.79) and (365.79,134) .. (368,134) .. controls (370.21,134) and (372,135.79) .. (372,138) .. controls (372,140.21) and (370.21,142) .. (368,142) .. controls (365.79,142) and (364,140.21) .. (364,138) -- cycle ;
\draw  [fill={rgb, 255:red, 0; green, 0; blue, 0 }  ,fill opacity=1 ] (390,138) .. controls (390,135.79) and (391.79,134) .. (394,134) .. controls (396.21,134) and (398,135.79) .. (398,138) .. controls (398,140.21) and (396.21,142) .. (394,142) .. controls (391.79,142) and (390,140.21) .. (390,138) -- cycle ;
\draw  [fill={rgb, 255:red, 0; green, 0; blue, 0 }  ,fill opacity=1 ] (375,147) .. controls (375,144.79) and (376.79,143) .. (379,143) .. controls (381.21,143) and (383,144.79) .. (383,147) .. controls (383,149.21) and (381.21,151) .. (379,151) .. controls (376.79,151) and (375,149.21) .. (375,147) -- cycle ;
\draw  [fill={rgb, 255:red, 0; green, 0; blue, 0 }  ,fill opacity=1 ] (405,149) .. controls (405,146.79) and (406.79,145) .. (409,145) .. controls (411.21,145) and (413,146.79) .. (413,149) .. controls (413,151.21) and (411.21,153) .. (409,153) .. controls (406.79,153) and (405,151.21) .. (405,149) -- cycle ;
\draw    (368,138) -- (379,147) ;
\draw    (409,149) -- (379,147) ;
\draw    (379,147) -- (394,138) ;
\draw    (409,149) -- (394,138) ;
\draw  [dash pattern={on 4.5pt off 4.5pt}]  (408,49.8) .. controls (447.4,20.25) and (447.02,82.87) .. (487.14,53.22) ;
\draw [shift={(489,51.8)}, rotate = 141.84] [fill={rgb, 255:red, 0; green, 0; blue, 0 }  ][line width=0.08]  [draw opacity=0] (8.93,-4.29) -- (0,0) -- (8.93,4.29) -- cycle    ;
\draw   (494,30) -- (605,30) -- (605,80.8) -- (494,80.8) -- cycle ;
\draw    (606,55.8) .. controls (645.2,26.4) and (622.93,87.27) .. (659.68,61.49) ;
\draw [shift={(662,59.8)}, rotate = 143.13] [fill={rgb, 255:red, 0; green, 0; blue, 0 }  ][line width=0.08]  [draw opacity=0] (8.93,-4.29) -- (0,0) -- (8.93,4.29) -- cycle    ;
\draw   (497,147.4) .. controls (497,138.78) and (504.16,131.8) .. (513,131.8) .. controls (521.84,131.8) and (529,138.78) .. (529,147.4) .. controls (529,156.01) and (521.84,163) .. (513,163) .. controls (504.16,163) and (497,156.01) .. (497,147.4) -- cycle ; \draw   (505.96,142.09) .. controls (505.96,141.23) and (506.68,140.53) .. (507.56,140.53) .. controls (508.44,140.53) and (509.16,141.23) .. (509.16,142.09) .. controls (509.16,142.96) and (508.44,143.65) .. (507.56,143.65) .. controls (506.68,143.65) and (505.96,142.96) .. (505.96,142.09) -- cycle ; \draw   (516.84,142.09) .. controls (516.84,141.23) and (517.56,140.53) .. (518.44,140.53) .. controls (519.32,140.53) and (520.04,141.23) .. (520.04,142.09) .. controls (520.04,142.96) and (519.32,143.65) .. (518.44,143.65) .. controls (517.56,143.65) and (516.84,142.96) .. (516.84,142.09) -- cycle ; \draw   (505,153.64) .. controls (510.33,157.8) and (515.67,157.8) .. (521,153.64) ;
\draw    (493,146.8) .. controls (445.72,189.15) and (479.94,112.17) .. (426.5,146.17) ;
\draw [shift={(424,147.8)}, rotate = 326.31] [fill={rgb, 255:red, 0; green, 0; blue, 0 }  ][line width=0.08]  [draw opacity=0] (8.93,-4.29) -- (0,0) -- (8.93,4.29) -- cycle    ;

\draw (10,10) node [anchor=north west][inner sep=0.75pt]   [align=left] {LTL property};
\draw (47,25) node [anchor=north west][inner sep=0.75pt]   [align=left] {$\displaystyle \psi $};
\draw (22,68) node [anchor=north west][inner sep=0.75pt]   [align=left] {Alphabet};
\draw (21,85) node [anchor=north west][inner sep=0.75pt]   [align=left] {$\displaystyle [ p,\ q,\ \ldots ]$};
\draw (22,129) node [anchor=north west][inner sep=0.75pt]   [align=left] {Truth valuee};
\draw (24,150) node [anchor=north west][inner sep=0.75pt]   [align=left] {$\displaystyle \top ,\ \bot ,\ \unk$};
\draw (238,90.9) node  [font=\large] [align=left] {Generator};
\draw (387.09,84.5) node   [align=left] {.hoa};
\draw (389.09,177.5) node   [align=left] {.gv};
\draw (549.5,55.4) node   [align=center] {Third-party\\library};
\draw (666,50) node [anchor=north west][inner sep=0.75pt]   [align=left] {$\displaystyle \ldots $};
\draw (501,170) node [anchor=north west][inner sep=0.75pt]   [align=left] {user};

\end{tikzpicture}

}
\caption{Overview of the tool.
}
\label{fig:tool}
\end{figure}
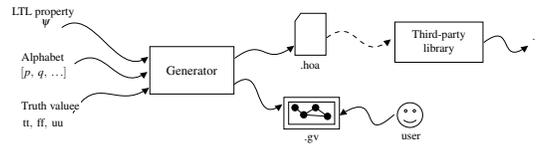

\new{Figure \ref{fig:tool} provides an overview of \ntool. The tool begins by parsing the user's input, which consists of three essential arguments. 
The first argument is the LTL property $\psi$ of interest, serving as a guide for generating the GNBA. The second argument represents the alphabet of $\psi$ and informs \ntool about the atomic propositions to consider when constructing the automaton. Since the automaton explicitly specifies the atomic propositions associated with its transitions, it is crucial to identify the relevant events of interest.
The third argument specifies the truth value against which the LTL formulas are verified. Following the approach proposed in~\cite{BelardinelliM20}, \ntool offers support for generating three different GNBA versions. To elaborate further, if $\top$ (representing satisfaction), $\bot$ (representing violation), or $\unk$ (representing neither satisfaction nor violation) is provided as the third argument, \ntool generates the respective GNBA, denoted as $A_{\psi,\top}$, $A_{\psi,\bot}$, or $A_{\psi,\unk}$, recognizing traces that satisfy, violate, or neither satisfy nor violate $\psi$.
}
\new{
\ntool produces two distinct output files. The first file, primarily graphical, contains the GNBA description in the DOT graph description language. The choice of DOT format stems from its widespread usage (supported by many programming languages) and its native compatibility with Graphviz\footnote{\url{https://graphviz.org/}}. 
}
\new{
The second file generated by \ntool adheres to the HOA (Hanoi Omega-Automata) format\footnote{\url{http://adl.github.io/hoaf/}}, a machine-readable format. This format enjoys support from well-known automata-based libraries, including Spot~\cite{DBLP:conf/atva/Duret-LutzLFMRX16} and LTL3BA~\cite{babiak2012ltl}\footnote{LTL3BA operates on two-valued automata, but its output is defined using three truth values, allowing it to effectively handle run-time verification scenarios. For a comprehensive examination of the distinctions between "undefined" and "unknown" truth values in the context of run-time verification, you can find more detailed information in \cite{DBLP:conf/sefm/FerrandoM22}.
}. This choice enhances compatibility with third-party tools, promoting the broader utility of the GNBA generated by \ntool. 
It is important to note that while \ntool operates independently, it seamlessly integrates with existing automata-based tools, ensuring a smooth transition for users and enabling further advancements and applications of the GNBA it produces.
}

\paragraph{Technical details.}

We go further into the detail of the implementation.
%
First of all, \ntool has been implemented in Java (version 17). The
resulting runnable jar can be directly used off-the-shelf. 


\ntool is divided into three components: input handler, automaton generator, and output handler.

\textit{Input handler.}
{
\ntool handles three input data: the LTL property $\psi$, the alphabet, and the truth value.
While the handling of the second and third arguments is straightforward, the first argument requires a bit more of work. Specifically, a parser has been implemented to parse LTL formulas
using Antlr\footnote{\url{https://www.antlr.org}}, which is directly supported in Java. The resulting visitor for the LTL grammar is not only used to parse the LTL property $\psi$ given in input, but it is also used to extract the corresponding closure $cl(\psi)$.
}

\textit{Automaton generator.}
{ After the LTL property $\psi$ given in input has been successfully
parsed, and the resulting closure $cl(\psi)$ has been generated, the
tool proceeds with the generation of the GNBA. The corresponding Java
object, instantiation of the custom $Automaton$ class, is generated
and stored in memory. Inside such object all information about states
and transitions, along with details on the initial and
accepting states, are stored. In particular, the set of
initial states is determined by the last input given to \ntool. If
the user desires to produce a GNBA to recognise the traces which
satisfy $\psi$, then the initial states in the automaton are all the
states containing $\psi$.
Note that, this is possible because the elementary subsets of
$cl(\psi)$ which determine the automaton's states are not necessarily
maximal, differently form the standard automaton
construction. Interestingly, the accepting states
in all three cases are the same.  }

\textit{Output handler.}
{ Once the GNBA has been generated and the corresponding Java object
is stored, \ntool moves forward to produce the resulting DOT and HOA
output files. Both files are generated using two different custom
methods of the $Automaton$ class. Such methods pass through all
states/transitions, and port these data in the wanted format.

\paragraph{Experiments.}

To show \ntool's scalability, we carried out some experiments
w.r.t. the size of the LTL formula given in
input. Figure~\ref{fig:exp} reports the results so obtained. As it is
easy to note, the results show \ntool is exponential w.r.t. the size
of LTL formula; where the size denotes the number of operators in the
formula (e.g., the LTL formula $XFp$ has size 2, while $Gp \land Fq$
has size 3). Note that, this was expected because the transformation
from LTL to GNBA is known to be exponential w.r.t. the size of the LTL
formula. So, \ntool extends the standard algorithm, but maintains the same
complexity.


\begin{figure}[ht]
    \centering
    \includegraphics[width=0.4\linewidth]{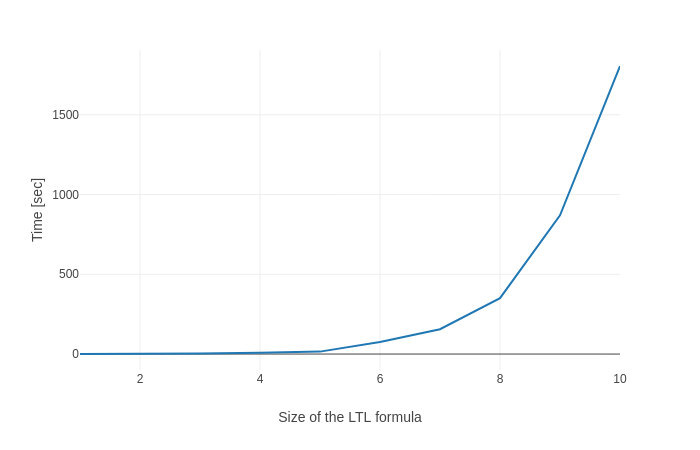}
    \caption{Experimental results.}
    \label{fig:exp}
\end{figure}

\vspace{-1em}

\section{Conclusions}\label{sec:conclusions}

In this paper, we have introduced a tool designed for generating automata from LTL formulas, interpreted within a three-valued semantics framework. To implement this tool, we have closely followed the automata construction methodology outlined in \cite{BelardinelliM20}.
Looking ahead, our future work entails extending the capabilities of our tool to accommodate more than three truth values. This extension would enable us to create a generator capable of handling multi-valued LTL formulas. Additionally, we envision applying the automata construction and its associated implementation in various domains related to multi-valued logics.
One such domain is Runtime Verification, where three-valued LTL also finds relevance. However, it is worth noting that the third value in Runtime Verification serves to maintain the impartiality of the monitor, while in our context, the third value signifies imperfect information about the system. As a result, our approach has the potential to address scenarios involving imperfect information, similar to the approach presented in~\cite{DBLP:conf/sefm/FerrandoM22}.
\new{Unfortunately, due to space constraints and the paper's primary focus, a comparative analysis with other tools has not been included.}

\bibliographystyle{eptcs}
\bibliography{generic}
\end{document}